\documentclass[aps,pra,twocolumn,groupedaddress,floatfix,superscriptaddress]{revtex4-1}

\usepackage{amsmath}
\usepackage{amssymb}
\usepackage{graphicx}
\usepackage{float}
\usepackage[english]{babel}
\usepackage{dsfont}
\usepackage{epstopdf}
\usepackage{soul}
\usepackage{color}
\usepackage{braket}

\usepackage[colorlinks=true,
            linkcolor=magenta,
        urlcolor=blue,
    citecolor=blue]{hyperref}

\usepackage[normalem]{ulem}

\definecolor{Nathanblue}{rgb}{0.12,0.24,0.40}

\def\be{\begin{equation}}
\def\ee{\end{equation}} 
\def\bea{\begin{eqnarray}}
\def\eea{\end{eqnarray}} 
\def\ba{\begin{array}} 
\def\ea{\end{array}}
\def\pa{\partial}

\def\Om{\Omega}

\def\nn{\nonumber}
\def\ket{\rangle}
\def\bra{\langle}
\def\b{\mathbf}

\def\f{\frac}

\def\ra{\rightarrow}
\def\rm{\mathrm}

\newcommand{\mv}[1]{\langle #1\rangle}

\emergencystretch=\maxdimen
\begin{document}
\title{Non-Abelian Bloch oscillations in higher-order topological insulators}
\author{M. Di Liberto}
\email{mar.diliberto@gmail.com}
\affiliation{Center for Nonlinear Phenomena and Complex Systems, Universit\'e Libre de Bruxelles, CP 231, Campus Plaine, B-1050 Brussels, Belgium}
\author{N. Goldman}
\affiliation{Center for Nonlinear Phenomena and Complex Systems, Universit\'e Libre de Bruxelles, CP 231, Campus Plaine, B-1050 Brussels, Belgium}
\author{G. Palumbo}
\affiliation{Center for Nonlinear Phenomena and Complex Systems, Universit\'e Libre de Bruxelles, CP 231, Campus Plaine, B-1050 Brussels, Belgium}

\maketitle

\textbf{
Bloch oscillations (BOs) are a fundamental phenomenon by which a wave packet undergoes a periodic motion in a lattice when subjected to an external force. Observed in a wide range of synthetic lattice systems, BOs are intrinsically related to the geometric and topological properties of the underlying band structure. This has established BOs as a prominent tool for the detection of Berry phase effects, including those described by non-Abelian gauge fields. In this work, we unveil a unique topological effect that manifests in the BOs of higher-order topological insulators through the interplay of non-Abelian Berry curvature and quantized Wilson loops. It is characterized by an oscillating Hall drift that is synchronized with a topologically-protected inter-band beating and a multiplied Bloch period. We elucidate that the origin of this synchronization mechanism relies on the periodic quantum dynamics of Wannier centers. Our work paves the way to the experimental detection of non-Abelian topological properties in synthetic matter through the measurement of Berry phases and center-of-mass displacements.
}

\emph{Introduction.} 
The quest for topological quantization laws has been a central theme in the exploration of topological quantum matter~\cite{Kane2010,Qi2011}, which originated from the discovery of the quantum Hall effect \cite{Klitzing1980, Girvin1999}. In the last decade, the development of topological materials has led to the observation of fascinating quantized effects, including the half-integer quantum Hall effect~\cite{Xu2014} and the quantization of Faraday and Kerr rotations~\cite{Wu2016} in topological insulators \cite{Fu2007, Qi2008}, as well as half-integer thermal Hall conductance in spin liquids~\cite{Kasahara2018} and quantum-Hall states~\cite{Banerjee2018}. In parallel, the engineering of synthetic topological systems has allowed for the realization of quantized pumps~\cite{Lohse2016,Nakajima2016,Lu2016,Lohse2018}, and revealed quantized Hall drifts~\cite{Aidelsburger2015, Genkina2019, Chalopin2020}, circular dichroism~\cite{Asteria2019}, and linking numbers~\cite{Tarnowski2019}.

In this context, Bloch oscillations (BOs) \cite{Waschke1993, BenDahan1996, Pertsch2002, Corrielli2013, Block2014, Meinert2017} have emerged as a powerful tool for the detection of geometric and topological properties in synthetic lattice systems~\cite{Choi1994,Price2012,Liu2013,Grusdt2014,Wang2016,Ramasesh2017,Flurin2017,Zheng2017}, hence providing access to quantized observables. Indeed, transporting a wave packet across the Brillouin zone can be used to explore various geometric features of Bloch bands, including the local Berry curvature~\cite{Price2012} and the Wilson loop of non-Abelian connections~\cite{Grusdt2014}. This strategy has been exploited to extract the Berry phase~\cite{Atala2013, Duca2015}, the Berry curvature~\cite{Jotzu2014,Wimmer2017}, the Chern number~\cite{Aidelsburger2015, Genkina2019, Chalopin2020}, and quantized Wilson loops~\cite{Li2016} in ultracold matter and photonics. 

\begin{figure}[!b]
\center
\includegraphics[width=0.95\columnwidth]{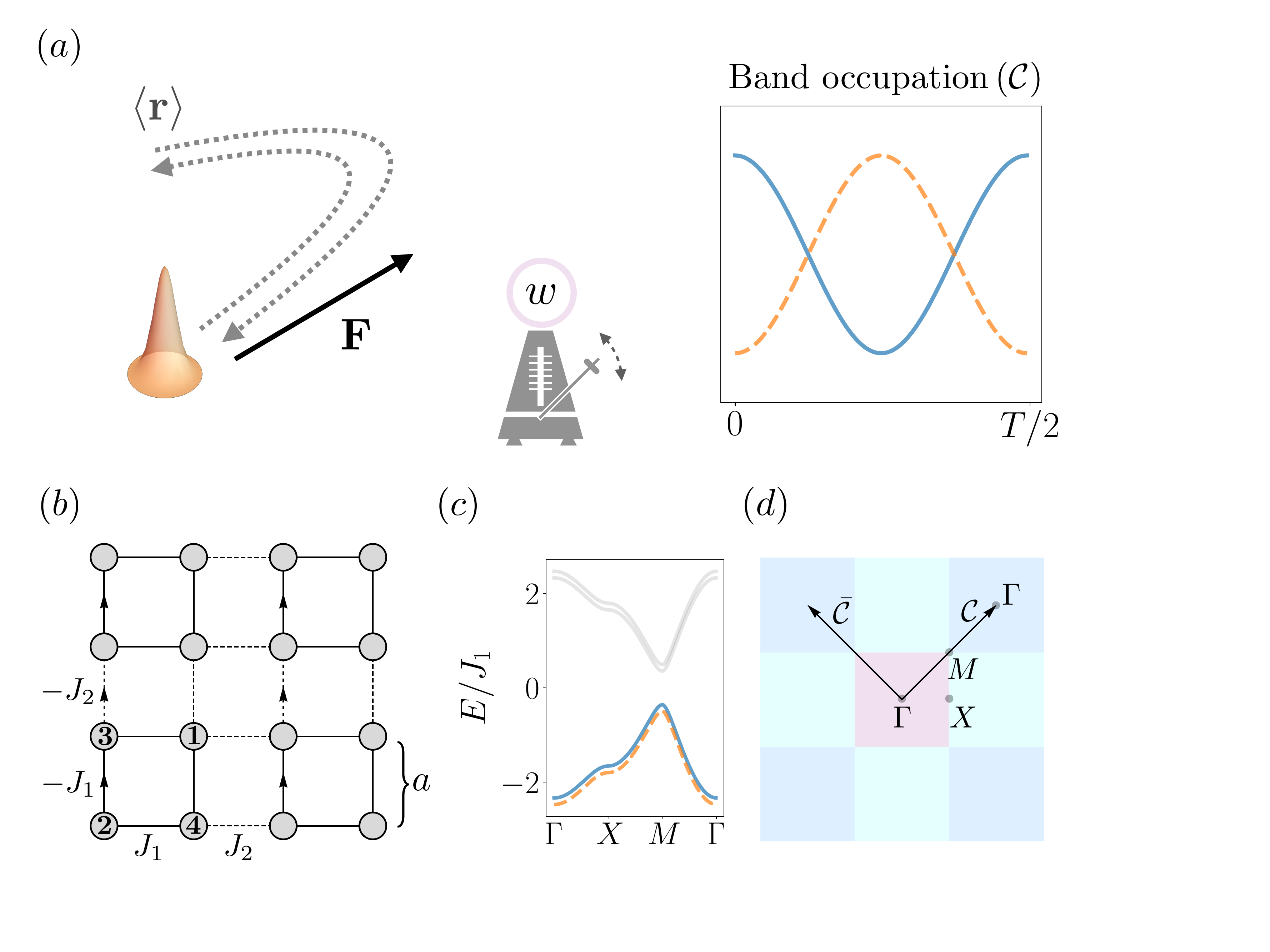}

\caption{Schematics of the non-Abelian topological BOs. (a) A gaussian wavepacket experiences a sign-changing Hall drift under the applied force $\b F$, while displaying a synchronized beating within two occupied bands. This synchronized effect, represented by the metronome, is topologically protected by the winding number $w$.  (b) BBH model:~a square lattice with $\pi$ flux and staggered hopping amplitudes $J_1$ and $J_2$. Vertical bonds with arrows correspond to a Peierls phase $\pi$ in the hopping amplitudes. (c) Band structure of the model. Each band is two-fold degenerate. (d) Brillouin zones and paths $\mathcal C$ and $\bar{\mathcal C}$ exhibiting topological BOs. 
}
\label{fig:model}
\end{figure}

The Wilson-loop measurement of Ref.~\cite{Li2016} highlighted a fundamental relation between two intriguing properties of multi-band systems: the quantization of Wilson loops, a topological property related to the Wilczek-Zee connection~\cite{Wilczek1984}, and the existence of ``multiple Bloch oscillations", which are characterized by a multiplied Bloch period~\cite{Li2016,Zheng2017,Zhang2017,Lang2017,Yan2019,Li2018,Li2019,Anderson2020}. The effect investigated in Ref.~\cite{Li2016} was eventually identified as an instance of ``topological Bloch oscillations", whose general framework was proposed in Ref.~\cite{Holler2018} based on the space groups of crystals and its implications on the quantization of geometric quantities (Zak phases differences). The Bloch period multiplier appears as a topological invariant, protected by crystalline symmetries, thus making multiple Bloch oscillations genuinely topological. Furthermore, when a Wannier representation of the bands is possible, Zak phases correspond to the positions of charges within the unit cell, namely the Wannier centers. As a consequence, a Zak-Wannier duality allows to connect BOs to the relative phases acquired by charges within a classical point-charge picture.  More recently, BOs displaying topologically-protected sub-oscillations have also been found in periodically driven systems in the context of quantum walks~\cite{Upreti2020}.

In this work, we identify a distinct topological effect that manifests in the BOs of higher-order topological insulators (HOTIs). These newly-discovered systems belong to the family of topological crystalline insulators~\cite{Fu2011, Fang2012, Hsieh2012, Morimoto2013, Slager2013, Shiozaki2014, Bernevig2016,Kruthoff2017}, \emph{i.e.}~gapped quantum systems characterized by crystal symmetries; they are characterized by quantized multipole moments in the bulk and unusual topologically-protected states (\emph{e.g.}~corner or hinge modes) on their boundaries; see Refs.~\cite{Benalcazar_Science, Benalcazar_PRB, Neupert2018,Brouwer2017,Song2017, Ezawa2018, Queiroz2019, Imhof2018, Serra-Garcia2018, Mittal2019, Chen2019, Kempkes2019, Zhang2019, Dutt2020, He2020}. Considering the prototypical Benalcazar-Bernevig-Hughes (BBH) model~\cite{Benalcazar_Science, Benalcazar_PRB}, we unveil a phenomenon by which multiple BOs take the form of an oscillating Hall drift, accompanied with a synchronized inter-band beating, for special directions of the applied force, as summarized in Fig.~\ref{fig:model}(a). While the Hall motion is attributed to the finite non-Abelian Berry curvature of the degenerate band structure, the inter-band beating captured by the Wilson loop is shown to be topologically protected by winding numbers. The synchronization of real-space motion and inter-band dynamics is elucidated through a quantum Rabi oscillation of Wannier centers. Finally, we observe that detached helical edge states are present on specific boundaries, compatible with the special symmetry axes associated with the topological BOs. A topological transition signaled by the sign change of the identified winding numbers and by the corresponding appearence/disapperance of these states is identified. 

Overall, our results demonstrate the rich interplay of non-Abelian gauge structures and winding numbers in the topological Bloch oscillations of HOTI's, but also establish Bloch oscillations as a powerful probe for non-Abelian topological properties in quantum matter.

\emph{Model and symmetries.} 
We consider the BBH model, as introduced in Ref.~\cite{Benalcazar_Science, Benalcazar_PRB}. It consists of a square lattice with alternating hopping amplitudes $J_1$ and $J_2$ in the two spatial directions and a $\pi$ flux per plaquette, as depicted in Fig.~\ref{fig:model}(b). We have introduced the flux by Peierls phases on the vertical links but other conventions can be used without affecting the results of this work.
The model is represented by a chiral-symmetric Hamiltonian of the form 
\be
\label{eq:dirac}
\hat H(\b k) = \sum_{i=1}^4 d_i (\b k) \Gamma^i\,,
\ee
where the $4\times 4$ Dirac matrices are written in the chiral basis $\Gamma^i \!=\! -\sigma_2\otimes\sigma_i$ for $i=1,\dots,3$ and $\Gamma^4\!=\!\sigma_1\otimes\mathcal I$. This model has two-fold degenerate energy bands  $E(\b k) \!=\! \pm \epsilon(\b k)$ with $\epsilon(\b k) \!=\! \sqrt{|\b d(\b k)|^2}$. The eigenfunctions of the lowest two bands read $ |u^1_{\b k}\ket = \f{1}{\sqrt{2}\epsilon} \left( d_1-id_2, -d_3 - id_4, 0 ,i \epsilon \right)^T$ and $ |u^2_{\b k}\ket \!=\! \f{1}{\sqrt{2}\epsilon} \left( d_3-id_4, d_1 + id_2, i \epsilon ,0 \right)^T$.

The full expressions of the $d_i(\b k)$'s for the \emph{isotropic} BBH model is $d_1(\b k)\!=\!(J_1\!-\!J_2) \sin(k_y/2)$, $d_2(\b k)\!=\!-(J_1\!+\!J_2) \cos(k_y/2)$, $d_3(\b k)\!=\!(J_1\!-\!J_2) \sin(k_x/2)$ and $d_4(\b k)\!=\!-(J_1\!+\!J_2) \cos(k_x/2)$ and the corresponding energy dispersion is displayed in Fig.~\ref{fig:model}(c). Here, we take the periodicity $d\!=\!2a\!=\!1$, where $a$ is the lattice spacing. The ordering of the unit cell sites chosen to represent the model in Eq.~\eqref{eq:dirac} is indicated in Fig.~\ref{fig:model}(b). Notice that the chosen basis takes into account the geometric shape of the unit cell. As a consequence, the Hamiltonian is not Bloch invariant, namely $H(\b k + \b G) \!\neq\! H(\b k)$, with $\b G$ a reciprocal lattice vector. 

The presence of time-reversal symmetry $\hat T$, with $\hat T^2=1$ and chiral symmetry $\hat S$, represented by $\Gamma^0 = \sigma_3 \,\otimes \, \mathcal I$, sets the model into the BDI class \cite{Kane2010}. Moreover, several crystalline symmetries are also present: two non-commuting mirror symmetries with respect to the $x$ and $y$ axis, namely $\hat M_x=\sigma_1\otimes\sigma_3$ and $\hat M_y=\sigma_1\otimes\sigma_1$, respectively; and a $\pi/2$ rotation symmetry $C_4= \left(\begin{smallmatrix} 0 & \mathcal I \\ -i\sigma_2 & 0 \end{smallmatrix}\right)$. The two mirror symmetries guarantee that inversion ($C_2$) is also a symmetry of the model, with $\hat C_2 = \hat M_x \hat M_y$. Moreover, the presence of mirror and rotation symmetries allows us to define a pair of mirror symmetries with respect to the diagonal axes of the lattice, $\hat M_{xy} = \hat M_y \hat C_4$ and $\hat M_{x\bar y}=-\hat M_x \hat C_4 $. This is one of the central ingredients allowing for topologically-protected BOs, as we will show below. 

Due to the two non-commuting mirror symmetries $\hat M_x$ and $\hat M_y$, the BBH model is a quadrupole insulator that has a quantized quadrupole moment in the bulk, vanishing bulk polarization and corner charges \cite{Benalcazar_Science}. The non-commutation of the mirror symmetries also provides a non-vanishing non-Abelian Berry curvature $\Omega_{xy}(\b k) = \pa_{k_x} A_y - \pa_{k_y} A_x -i [ A_x,  A_y]$ of the two-fold degenerate lowest (or highest) bands \cite{Marzari1997, Nagaosa2010}, where $\b A$ is the non-Abelian Berry connection \cite{Wilczek1984}. However, the total Chern number of the degenerate bands remains zero due to time-reversal. It is then possible to define Wannier functions $|\nu^{\alpha}_{x,k_y}\ket$ and $|\nu^{\alpha}_{y,k_x}\ket$, with $\alpha=1,2$ numbering the bands below the energy gap, which are eigenstates of the position operators $\hat P\hat x \hat P$ and $\hat P\hat y \hat P$ projected onto the lowest two bands, respectively \cite{Benalcazar_Science, Benalcazar_PRB}. The non-commutation of the mirror symmetries (and therefore of the projected position operators) forces the use of hybrid Wannier functions, namely Wannier states that can only be maximally localized in one direction \cite{Marzari2012, Vanderbilt2014}. Furthermore, it provides a necessary condition to have gapped Wannier bands, namely Wannier centers that are displaced from each other at every momentum $k_x$ or $k_y$. In Ref.~\cite{Benalcazar_Science, Benalcazar_PRB}, the Wannier gap has been exploited to define a winding of the Wannier states (nested Wilson loop) as a condition to have a quantized quadrupole moment in the bulk, which can be revealed from the Wannier-Stark spectrum \cite{Poddubny2019}.

\emph{Winding numbers.}
We now prove that a non-trivial topological structure captured by novel winding numbers characterizes the BBH model along the diagonal paths of the Brillouin zone, $\mathcal C$ and $\bar{\mathcal C}$, which are shown in Fig.~\ref{fig:model}(d). In order to emphasize the generality of these results, we hereby consider a generic Dirac-like model [Eq.~\eqref{eq:dirac}] without specifying the components of the $\b d(\b k)$ vector. We assume that all previously discussed symmetries are satisfied with the additional constraint that each $d_i$ function only depends on one component of the momentum $\b k$, namely we assume that $d_1 = d_1(k_y)$, $d_2 = d_2(k_y)$, $d_3 = d_3(k_x)$, $d_4 = d_4(k_x)$. Such constraint is satisfied by the BBH model. Mirror symmetries impose that $d_1$ and $d_3$ are odd functions whereas $d_2$ and $d_4$ are even. We then find that along $\mathcal C$ (\emph{i.e.} for $k\!=\!k_x\!=\!k_y$), the diagonal mirror symmetry represented by the operator $\hat M_{xy}$ requires $d_3(k)=d_1(k)$ and $d_4(k)=d_2(k)$, while along $\bar{\mathcal C}$ (\emph{i.e.} for $k\!=\!k_x\!=\!-k_y$), the symmetry operator $\hat M_{x\bar{y}}$ requires $d_3(k)=-d_1(k)$ and $d_4(k)=d_2(k)$. We then conclude that only two components of $\b d$ are independent and we therefore define the vector $\tilde{\b d}(k) \equiv (d_1(k),d_2(k))$. After writing the Hamiltonian in its chiral representation $\hat H(\b k) = \left( \begin{smallmatrix} 0 & Q(\b k) \\ Q(\b k)^\dag & 0 \end{smallmatrix} \right)$, where $Q(\b k) = d_4(\b k) \mathcal I + i d_i(\b k) \sigma^i$, we obtain the following result 
\begin{align}
w_{ _{\mathcal{C} ( {\bar{\mathcal C}} ) } }  &\equiv \f{i}{2\pi} \int_{\mathcal C} \mathrm d \b k\, \cdot \mathrm{Tr} \left[ Q(\b k)^{-1} \sigma_{3 (1)} \, \pa_{\b k} Q(\b k) \right] \nn \\
& = -\f 1 \pi \int_0^{2\pi} \mathrm d k \, \varepsilon^{ij} \f{\tilde d_i \pa_k \tilde d_j }{|\tilde{\b d}|^2} = \mathrm{sign} (J_1^2-J_2^2) \,,
\end{align}
where $\varepsilon^{12}=-\varepsilon^{21}=1$, and where we used the $\tilde{\b d}$ vector of the BBH model in the last step.

We therefore conclude that the quantities $w_{_\mathcal C}$ and $w_{_{\bar{\mathcal C}}}$ count how many times the vector $\tilde{\b d}$ winds over the closed paths $\mathcal C$ and ${\bar{\mathcal C}}$, respectively. The quantized windings $w_{_\mathcal C}$ and $w_{_{\bar{\mathcal C}}}$ are here protected by the crystalline symmetries $\hat M_x$, $\hat M_y$ and $\hat C_4$, as shown in the Supplemental Material. 
These symmetries also imply that the two topological invariants are not independent. We point out that similar winding numbers have been introduced in chiral-symmetric one-dimensional topological superconductors \cite{Dumitrescu2013, Daido2019}. 

Finally, the sign change of the winding numbers at the gap closing point $J_1=J_2$ signals a phase transition. We will show below that the transition corresponds to the appearance of detached helical edge states. Let us now focus on the BOs of the BBH model and the role played by the quantized winding numbers discussed above.

\begin{figure}[!t]
\center
\includegraphics[width=0.95\columnwidth]{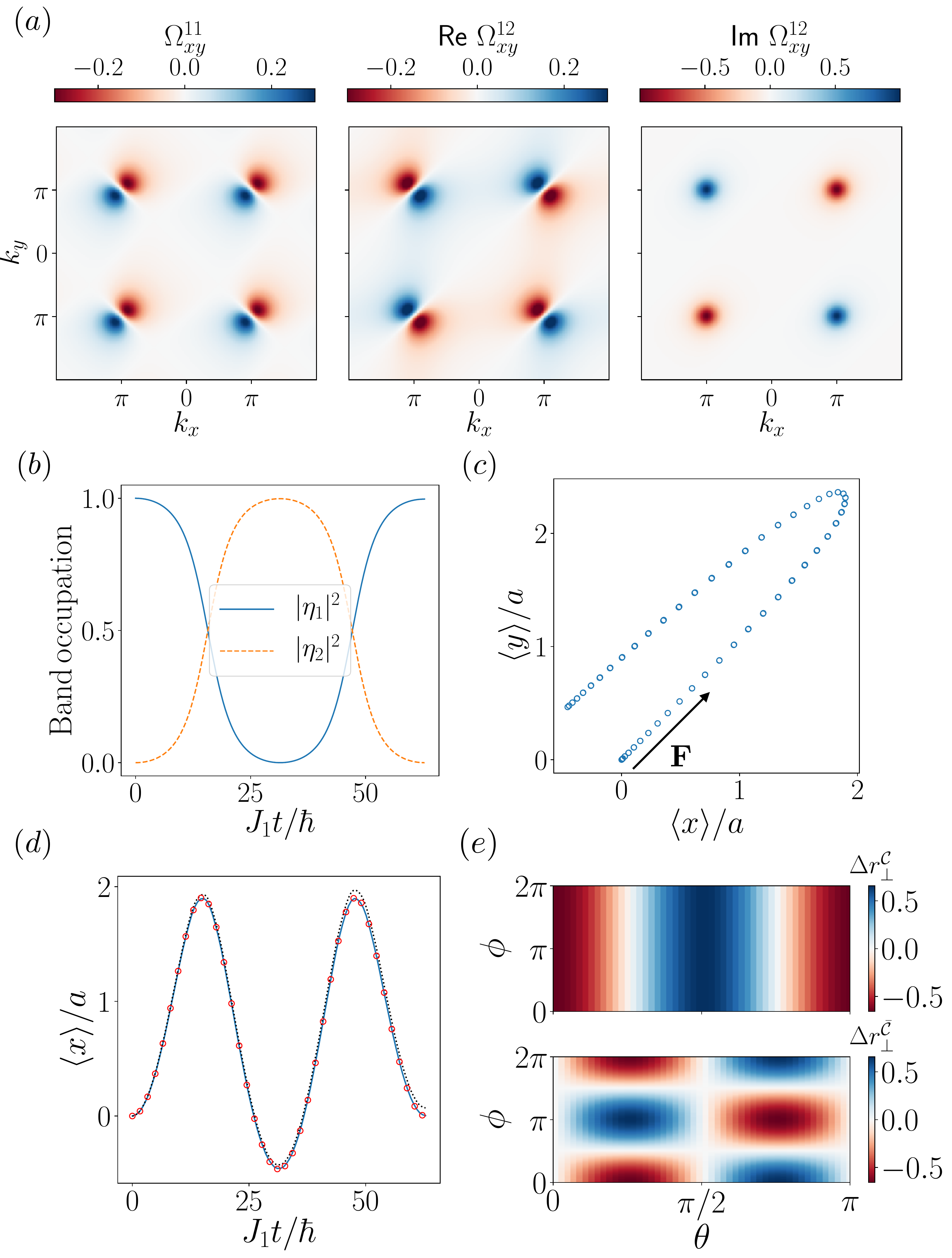}

\caption{Time dynamics of multiple BOs. (a) Non-Abelian Berry curvature profile for the BBH model. (b) Band occupation dynamics along the $\mathcal C$ path for $\theta(0) = 0$ and $\phi(0) = 0$. Here $J_2 = 0.3 J_1$ and $|F|=0.2\sqrt 2 J_1$. (c) Real space wavepacket trajectory. (d) Comparison of the wavepacket $\mv x$ position for (solid line) numerical real space evolution of a wavepacket with width $\sigma = 0.15 a^{-1}$ and momentum grid-spacing $k_{\textrm{pts}} = 20$, (dashed line) exact evolution, (dots) semiclassical evolution. (e) Orthogonal displacement after one BO along the paths $\mathcal C$ and $\bar{\mathcal C}$.
}
\label{fig:blochosc}
\end{figure}

\emph{Topological Bloch oscillations: band-population dynamics.} 
We consider a wavepacket obtained as a superposition of the lowest two bands and centered at $\b k$, which we write as $| u_{\b k}(t) \ket = \eta_1(t) | u^1_{\b k} \ket + \eta_2(t) | u^2_{\b k} \ket$ with $\eta = (\eta_1, \eta_2)^T$
\footnote{Owing to the degeneracy of the states \unexpanded{$| u^{1,2}_{\mathbf k} \rangle$}, other parametrizations can be chosen. As shown in Supplemental Material, this gauge ambiguity can be removed by weakly breaking time-reversal symmetry, thus splitting the two states in energy.}.
Under an applied homogeneous and constant force $\b F$, which makes the crystal momentum change linearly in time, $\dot{\b k} = \b F$,
the bands occupation evolves according to \cite{Culcer2005,Shindou2005}
\be
\label{eq:etat}
\dot \eta =  -i \epsilon_{\b k} \eta + i \b F \cdot \b A \eta\,,
\ee
where the matrix elements of the Berry connection are defined as $A^{\alpha\beta}_i = i \mv{u^\alpha_{\b k} | \pa_{k_i} | u^\beta_{\b k}}$. Here, the force is assumed to be weak enough so that transitions to upper bands are neglected.

We can formally solve Eq.~\eqref{eq:etat} as $\eta(t) = \exp( -i \int_0^t \mathrm d t \, \epsilon_{\b k} )   W\, \eta(0)$. The Wilson line operator $W$ is defined as  $ W \!=\!  \mathcal T \exp ( i \int_0^t \mathrm d t \, \b F \cdot \b A ) \!=\!  \mathcal P \exp( i \int_{\b k_i}^{\b k_f} \b A \cdot \mathrm d \b k )$, where we have denoted as $\b k_i$ and $\b k_f$ the initial and final momenta of the BO, respectively.
For a closed path $\mathcal C_0$ with $\b k_f = \b k_i + \b G$, where $\b G$ is a reciprocal lattice vector, the bands population dynamics is determined by the Wilson loop matrix $W_{\mathcal C_0} = \mathcal P \exp(i \int_{\mathcal C_0}\b A \cdot \mathrm d \b k )$.

Importantly, the winding numbers $w_{_{\mathcal C (\bar{\mathcal C})}}$ that we have previously introduced appear in the Wilson loops defined along the diagonal paths $\mathcal C$ and $\bar{\mathcal C}$, as $W_{\mathcal C (\bar{\mathcal C})} = \exp ( i (2\pi/4) w_{_{\mathcal C (\bar{\mathcal C})}} \sigma_{1(3)} )$, with $w_{_{\mathcal{C}(\bar{\mathcal C})}}=\pm 1$.  From this, we obtain that BOs require four loops in momentum space in order to map the wavefunction back to itself, namely $[W_{\mathcal C (\bar{\mathcal C})}]^4=\mathcal I$. Notice that the degeneracy of the bands brings a trivial dynamical phase that does not influence the internal band-population dynamics.

According to the classification of topological BOs discussed in Ref.~\cite{Holler2018}, rotational symmetries $\hat C_n$ can quantize BOs with a force applied orthogonal to the rotational symmetry axis. This is (partially) the case here, with $\hat C_4$ providing period-four BOs. However, $\hat C_4$ symmetry alone is not sufficient to quantize the BOs. Additional symmetries, namely $\hat M_{x}$ and $\hat M_y$, are required in order to have a protected winding number along the paths $\mathcal C$ and $\bar{\mathcal C}$, see Supplemental Material. In the general framework presented in Ref.~\cite{Holler2018}, a ``Wannier-Zak" relation is demonstrated when mirror symmetries commute. As a consequence, the Zak phase winding that appears in the Wilson loop has a one-to-one correspondence with the position of the Wannier centers. This is well described by independently evolving point charges within a classical picture. In our case, such a direct correspondence is not possible due to the non-vanishing Berry curvature and we will see below that the physical consequences of this feature appear on the real-space motion of the wavepacket. 

\emph{Topological Bloch oscillations: Real-space dynamics.}
Let us now consider the real-space motion of the wavepacket's center of mass, which satisfies the following semiclassical equations \cite{Xiao2010,Culcer2005,Shindou2005}
\begin{align}
\label{eq:semicl}
\dot x &= \pa_{k_x} \epsilon_{\b k} - F_y \eta^\dag \Omega_{xy} \eta \,, \nn \\
\dot y &= \pa_{k_y} \epsilon_{\b k} + F_x \eta^\dag \Omega_{xy} \eta \,.
\end{align}
Here, $\Omega_{xy}$ denotes the SU(2) Berry curvature, whose components are shown in Fig.~\ref{fig:blochosc}(a) for the BBH model; they satisfy the following conditions: $\Omega^{11}_{xy} \!=\! -\Omega^{22}_{xy}$, $ \mathrm{Re}\, \Om^{12}_{xy} \!=\!  \mathrm{Re}\, \Om^{21}_{xy}$ and $\mathrm{Im} \,\Om^{12}_{xy} \!=\! - \mathrm{Im}\, \Om^{21}_{xy}$. One anticipates from the accumulation of Berry curvature near the $M$ point of the Brillouin zone (BZ) that the paths $\mathcal C$ and $\bar{\mathcal C}$ may display nontrivial features also in the real-space dynamics and not only in the band population beating discussed above. As we shall explain in detail below, the wavepacket experiences a transverse Hall drift that changes sign after each BO, thus bringing the center of mass position back to its initial point after two BOs. This behavior is synchronized and tightly connected with the band-population dynamics captured by the Wilson loop.

The two-fold degeneracy of the bands allows us to parametrize the evolving state $| u_{\b k}(t) \ket$ on the Bloch sphere as $\eta_1(t) = \cos \theta(t)$ and $\eta_2(t) = \sin\theta(t) e^{i\phi(t)}$. We can therefore rewrite the anomalous velocity as 
\begin{align}
\label{eq:anvel}
\eta^\dag \Omega_{xy} \eta =& \, (|\eta_1|^2 - |\eta_2|^2)\Omega^{11}_{xy} + \sin2\theta \cos\phi \, \mathrm{Re} \, \Om^{12}_{xy} \nn \\
& - \sin 2\theta \sin\phi \, \mathrm{Im}\, \Om^{12}_{xy} \,.
\end{align}
On the $\mathcal C$ path, the angle $\phi$ is a constant of motion, namely $\dot\phi = 0$. This means that the Bloch vector is confined to a meridian of the Bloch sphere. Moreover, since $\mathrm{Re} \, \Om^{12}_{xy} = 0$ on $\mathcal C$, only the first and the last term of Eq.~\eqref{eq:anvel} are relevant. After one BO, the two bands populations exchange, symmetrically with respect to the $M$ point, as displayed in Fig.~\ref{fig:blochosc}(b) [i.e.~$W_{\mathcal C}\propto \sigma_1$]. Let us consider the case with $\phi=0$ and $\theta(0) = 0$, where only the first term in Eq.~\eqref{eq:anvel} matters. The Berry curvature has a node and the band population starts with $\eta_1(0)=1$. Near the $M$ point and before crossing it, the occupation of band $1$ is larger than the occupation of band $2$. The Berry curvature $\Omega^{11}_{xy}$ is positive and the Hall displacement in the $x$ direction is therefore negative (see the minus sign in the first of Eqs.~\eqref{eq:semicl}). Once the path has crossed the $M$ point, the occupations are flipped but so is also the sign of the Berry curvature, thus the Hall displacement continues with the same sign until the wavepacket reaches the $\Gamma$ point. Since the bands occupations have exchanged, a second BO will experience an opposite Hall drift and bring back the wavepacket to its initial position, as shown in Figs.~\ref{fig:blochosc}(c-d). From this, we obtain that the real-space motion is a witness of the non-Abelian band dynamics. For $\theta(0)\neq 0$, the off-diagonal component of the Berry curvature also contributes, and it fully suppresses the Hall displacement when $\theta(0) = \pi/4$ since the anomalous Hall velocity vanishes identically: the two bands are equally populated, no band exchange takes place and therefore the positive and negative deflections compensate each other.  

On the $\bar{\mathcal C}$ path, the Berry curvature has only off-diagonal components and the Hall dynamics is determined by the relative phase $\phi$, whereas $\theta$ is a constant of motion. In this case, the populations of the two bands do not exchange over time but the relative phase does by an angle $\pi$. The transverse displacement as a function of $\theta(0)$ and $\phi(0)$ for the two paths is shown in Fig.~\ref{fig:blochosc}(e). 

We have thus found that the center of mass of the wavepacket displays a period-two BO instead of a period-four one. This difference with respect to the Wilson loop analysis occurs because after two BOs, the wavefunction has picked up an overall phase $2\times (\pi/2)$, which does not appear in observables $\mv{\hat O}$, such as for the center of mass position.


\begin{figure}[!t]
\center
\includegraphics[width=0.95\columnwidth]{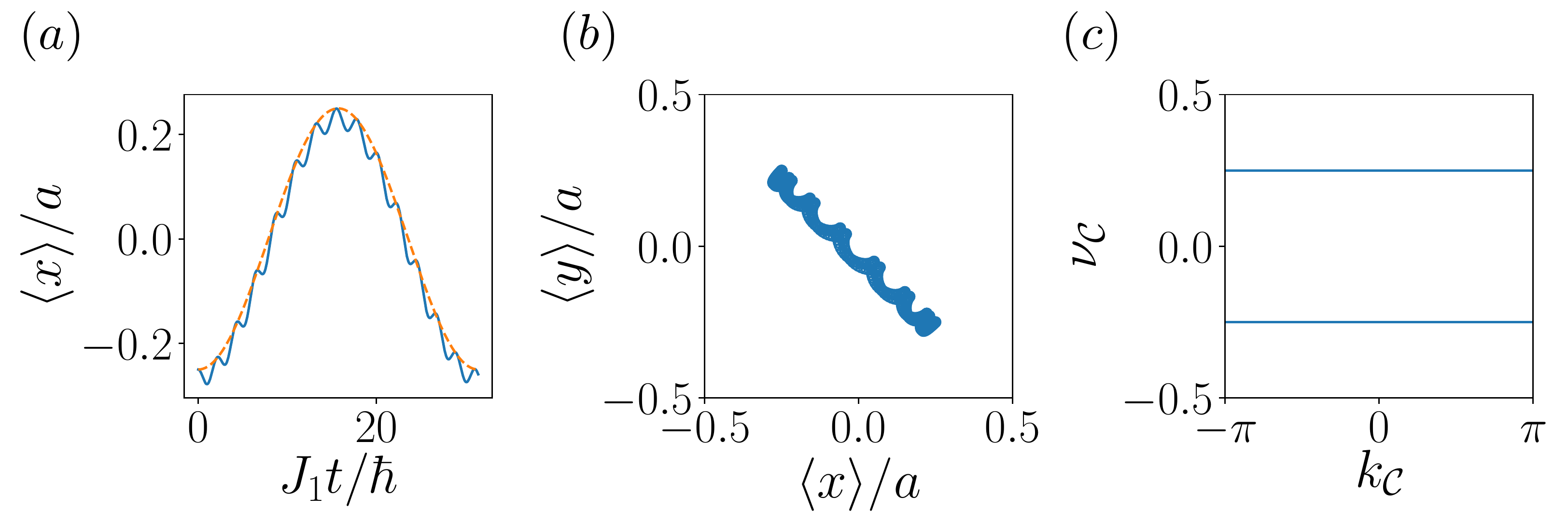}

\caption{Single plaquette dynamics. (a) Comparison between (solid line) the exact dynamics and (dashed line) the projected model. (b) Full real space evolution. (c) Hybrid Wannier centers obtained from diagonalizing the Wilson loops orthogonal to the $\mathcal C$ path away from the atomic limit for $J_1 = 0.3 J_2$. Here the force is aligned along the diagonal $F_x\!=\!F_y$.  
}
\label{fig:plaquette}
\end{figure}

\emph{Atomic limit: single-plaquette dynamics.} 
In order to elucidate the role of Wannier functions in the BOs analyzed here and the synchronization between real space and band-population dynamics, we consider the instructive atomic limit with $J_2 = 0$, where we can study the time dynamics of a single plaquette. The lowest energy eigenstates read $|u^1\ket  = (1/2, 1/2, 0, 1/\sqrt 2)^T$ and $|u^2\ket = (1/2, -1/2, 1/\sqrt 2, 0)^T$. 

We construct the position operator $\hat{\b r} \!=\! \sum_i (\b r_i \!-\! \b r_0) | \b r_i \ket \bra \b r_i |$ by setting the spatial origin at the plaquette center. We obtain the matrices $\hat x / a = \textrm{diag}(1/2, -1/2, -1/2, 1/2)$ and $\hat y / a = \textrm{diag}(1/2, -1/2, 1/2, -1/2)$. Let us now call $\hat P$ the projector operator on the states $|u^1\ket$ and $|u^2\ket$, from which we can construct the projected position operators  $\hat P \hat x \hat P\equiv \hat x_{_P} = \sum_{\alpha,\beta = 1,2} |u^\alpha \ket \mv{u^\alpha | \hat x | u^\beta} \bra u^\beta |$ and $\hat P \hat y \hat P\equiv \hat y_{_P} =  \sum_{\alpha,\beta = 1,2} |u^\alpha \ket \mv{u^\alpha | \hat y | u^\beta} \bra u^\beta |$. It follows that $[\hat x_{_P}, \hat y_{_P}] \neq 0$ whereas $\{ \hat x_{_P}, \hat y_{_P} \} = 0$. The eigenfunctions of the projected position operators are the Wannier functions $|\nu_{x,y}\ket$ and the corresponding eigenvalues are the Wannier centers \cite{Marzari1997, Alexandradinata2014, Benalcazar_PRB}, which read here $\nu_x=\nu_y=\pm a/4$. 

In the presence of an external tilt (or electric field), the perturbative Hamiltonian governing the dynamics for small values of the force $\b F$ reads 
\begin{align}
\label{eq:Rabi}
\hat H_F &= \b F \cdot \hat{\b r}_{_P} = F_x \, \hat x_{_P} + F_y \, \hat y_{_P} \nn\\
&= \f a 4 (F_x + F_y) \sigma_1 + \f a 4 (F_x-F_y) \sigma_3 \,.
\end{align}
The projected Hamiltonian reveals how the external force induces a quantum dynamics between the eigenstates of non-commuting position operators, in the form of a Rabi oscillation. This result is in sharp contrast with the classical point-charge picture introduced in Ref.~\cite{Holler2018}, which is valid for commuting position operators. We can now diagonalize $H_F$ and we find the spectrum $E = \pm Fa / 2\sqrt{2}$, with $F=\sqrt{F_x^2 + F_y^2}$\,. The Rabi period can be easily obtained as $T_R = 2\pi/(Fa/\sqrt 2)$. This solution is general and it does not depend on the direction of the force. Besides, we can always rotate the coordinate system in order to have one axis parallel to the force and one axis orthogonal to it, $ r^\|$ and $ r^\perp$, and reduce the Hamiltonian to $\hat H_F = F^\| \, \hat r^\|_{_P}$. Then, the corresponding time dynamics can be represented by the eigenstates of $\hat r^\perp_{_P}$, namely the Wannier functions obtained by diagonalizing $\hat r^\perp_{_P}$. As a consequence, we observe a transverse dynamics compared to the direction of the applied force $\b F$, as shown in Figs.~\ref{fig:plaquette}(a),(b). 

However, the Wannier centers dynamics is not directly connected to the BOs and its period does not have to be the same as the Rabi period of the Wannier centers. For example, let us consider a BO with $F_y = 0$. The periodicity of the BO occurs at the discrete times $T_B = 2\pi n/d F_x$, for $n\in \mathbb Z_+$ where $d=2a$. There is no solution that satisfies $T_B = T_R$. However, if we take $F_x = \pm F_y$ we find that $n=2$ provides $T_B = T_R$. Therefore, a force oriented along the diagonal axes allows to synchronize the Wannier centers dynamics with the BOs, whereas the other directions yield out-of-sync oscillations that does not bring the wavepacket back to its initial position at integer multiples of the fundamental Bloch period. 

Away from the atomic limit, we can still use Wannier functions as a complete basis to express the wavepacket. A direct calculation (see Fig.~\ref{fig:plaquette}(c)) shows that along the paths $\mathcal C$ and $\bar{\mathcal C}$, the Wannier centers remain gapped and their spectrum flat, namely they are equispaced along the entire path. We interpret this fact as a witness that the Wannier centers can be thought as oscillators with the same oscillation frequency (\emph{i.e.} displacement), as in the atomic limit represented by Eq.~\eqref{eq:Rabi}, thus keeping the same oscillatory motion while changing the momentum $k_{_\mathcal{C}}$ or $k_{_{\bar{\mathcal C}}}$. In conclusion, Wannier centers perform a quantum periodic dynamics where their transverse motion with respect to the applied force is periodic and synchronized with the BO period.


\begin{figure}[!t]
\center
\includegraphics[width=0.95\columnwidth]{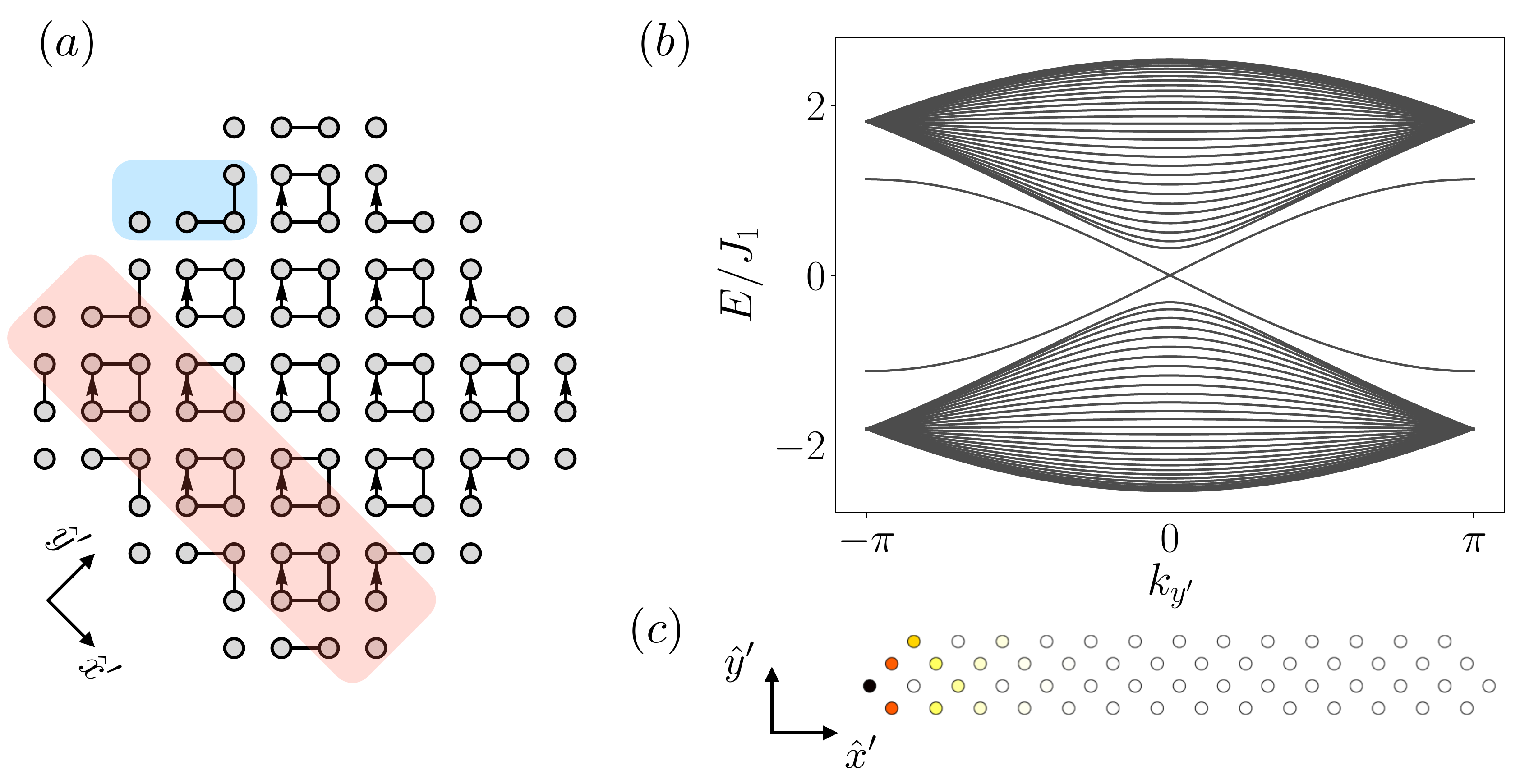}

\caption{Open lattice and edge states. (a) Open lattice in the atomic limit respecting mirror $M_{x,y}$ and $C_4$ symmetries. Highlighted in blue the edge sites displaying zero modes and in red the unit cell for the stripe geometry. (b) Energy spectrum obtained by imposing periodic boundary conditions along $\hat y'$ and $J_1 = 0.8 J_2$. Each edge hosts a pair of detached helical edge modes. (c) Positive energy edge state for $k_{y'} = k_\mathcal{C} = \pi/2$.
}
\label{fig:edge}
\end{figure}

\emph{Edge states.} 
The quantized winding numbers $w_{_{\mathcal C}}$ and $w_{_{\bar{\mathcal C}}}$, which we have previously identified along the paths $\mathcal C$ and $\bar{\mathcal C}$, indicate that a topological transition takes place when $J_1 = J_2$. Here, we show that an open system with edges along the diagonals $x \pm y$ displays detached helical  edge states. From Fig.~\ref{fig:edge}(a), we notice that near the atomic limit $J_2\ra 0$, the bulk has gapped states at energies $E_b\sim \pm\sqrt 2 J_1$. The edge displays disconnected single sites at energy $E_s\sim 0$ and trimers, with energies $E_{t1}\sim 0$ and $E_{t2} = \pm \sqrt 2 J_1$. Thus, a pair of zero energy ($E_s$ and $E_{t1}$) modes exists at the edge.

Near the gap closing point, $J_2 = (1+m)J_1$ with $|m|\ll 1$, we construct an effective continuum theory \cite{Zhang2009, Shen2012} for two (pseudo)-spins satisfying
\be
\left( 2 m + \f{1}{2} \pa_{x'}^2 \right) \sigma_2 \psi_{\uparrow,\downarrow}(x') = \pm \pa_{x'} \psi_{\uparrow,\downarrow}(x') \,.
\ee
These equations provide two independent zero-energy solutions
\begin{align}
\psi(x') & = 
\begin{pmatrix}
\chi_- \\
0
\end{pmatrix}
e^{-2x'} \left( e^{2mx'} - e^{-2mx'} \right)\,, \nn \\
\psi(x') & = 
\begin{pmatrix}
0 \\
\chi_+
\end{pmatrix}
e^{-2x'} \left( e^{2mx'} - e^{-2mx'} \right) \,,
\end{align}
where $\sigma_2 \chi_\eta = \eta \chi_\eta$ and $\eta = \pm 1$, which are localized at $x'=0$ and exist only for $m>0$, namely when $J_1<J_2$ (see Methods section). To compute the dispersion relation of the edge modes, it is convenient to consider a cylindrical geometry. In this case, we find that the edge modes become helical, see Fig.~\ref{fig:edge}(b). An example of such states is shown in Fig.~\ref{fig:edge}(c). 

\emph{Discussion and conclusions.} 
In this work, we have shown a new type of multiple Bloch oscillations that is connected to the quantum beating of Wannier centers and we have identified higher-order topological insulators as a model where this effect can be observed. By studying the BBH model, we have shown that the Wilson loop imposes period-four oscillations and the center-of-mass motion displays an anomalous Hall displacement over one period of oscillation. We have connected these features to the crystalline symmetries of the model and we have identified quantized winding numbers that protect the topological BOs. Moreover, we have shown that detached helical edge states emerge in an open system with the required symmetries.

Our results can be observed with cold atoms \cite{Goldman2016a, Cooper2019}, where flux engineering can be achieved through time-dependent protocols \cite{Aidelsburger2011, Aidelsburger2013} and where the staggered hopping amplitudes requires a bipartite lattice \cite{Atala2013, DiLiberto2014}. Interferometric and tomographic methods can be exploited to measure the Wilson loop winding \cite{Grusdt2014, Li2016, Sugawa2019} and real-space cloud imaging makes possible to measure the center-of-mass displacement \cite{Wintersperger2020}. A fundamental question concerns the preparation of the initial state, owing to the degenerate nature of the bands. As shown in Supplemental Material, the bands can be split by slightly breaking time-reversal symmetry. In this case, it is possible to prepare a non-degenerate Bose-Einstein condensate (BEC) at the $\Gamma$ point. When projected onto the eigenstates of the BBH model, this state is peaked at specific values of $\theta$ and $\phi$. One can then obtain the desired superposition of the two zero-momentum modes (the BEC and the gapped mode) by a coherent coupling through an external driving. The subsequent BOs require that the applied force has a magnitude that is larger than the band separation to effectively recover the band degeneracy during the BOs.

In the context of photonics, our results can be investigated by using optical waveguides \cite{Ozawa_Rev}, where it has been recently possible to realize synthetic $\pi$ flux \cite{Mukherjee2018, Kremer2020}. In this platform, the input laser profile can be inprinted in order to map the degenerate manifold of states at the $\Gamma$ point that are parametrized by the angles $\theta$ and $\phi$. It is then possible to reconstruct the Wilson loop dynamics by measuring the output field phase profile, whereas the Hall displacement is obtained from the spatial profile of the field intensity.

As a perspective of our work, it would be interesting to generalize our results to other two- and three-dimensional topological crystalline insulators and consider corrections to the semiclassical equations, \emph{e.g.} involving the quantum metric once an inhomogeneous electric field or a harmonic trap potential are introduced \cite{Bleu2018, Lapa2019}. Finally, given the role played by the initial state in the observation of the anomalous Hall displacement, BOs can be thought as a tool to witness the phenomenology of symmetry-broken condensates where the ground state degeneracy has been removed by interactions \cite{DiLiberto2020}. 

\emph{Acknowledgements.} 
We acknowledge M. Aidelsburger, W. Benalcazar, F. Grusdt, H. M. Price, G. Salerno for fruitful discussions. This work is supported by the ERC Starting Grant TopoCold, and the Fonds De La Recherche Scientifique (FRS-FNRS, Belgium).


\section*{Methods}

\textbf{Berry connection and curvature.}
For the BBH model introduced in the main text, the corresponding matrix elements of the non-Abelian Berry connection, defined as $A^{\alpha\beta}_i = i \mv{u^\alpha_{\b k} | \pa_{k_i} | u^\beta_{\b k}}$, read 
\begin{align}
A^{11}_x &= - A^{22}_x= -\f{J_1^2-J_2^2}{4\epsilon_{\b k}^2} \,, \\
A^{12}_x &= (A^{21}_x)^* = e^{-i\f{k_x+k_y}{2}}\f{(e^{ik_y}J_1+J_2)(J_1-e^{i k_x}J_2)}{4\epsilon_{\b k}^2} \,, \nn \\
A^{11}_y &= -A^{22}_y = \f{J_1^2-J_2^2}{4\epsilon_{\b k}^2} \, \nn\\
A^{12}_y & =(A^{21}_y)^* = e^{-i\f{k_x+k_y}{2}}\f{(e^{ik_y}J_1-J_2)(J_1+e^{i k_x}J_2)}{4\epsilon_{\b k}^2} \,. \nn
\end{align}
The SU(2) Berry curvature, defined as $\Omega_{xy}(\b k) = \pa_{k_x} A_y - \pa_{k_y} A_x -i [A_x, A_y]$, reads
\begin{align}
\Omega_{xy}^{11} &= - \Omega_{xy}^{22} = J_1 J_2 (J_1^2 - J_2^2) \f{\sin k_x + \sin k_y}{4\epsilon_{\b k}^4}\,, \\
\Omega_{xy}^{12} &= (\Omega_{xy}^{21})^* = - i (J_1^2-J_2^2) e^{-i \f{k_x+k_y}{2}}\f{ e^{ik_y} J_1^2 - e^{ik_x} J_2^2}{4\epsilon_{\b k}^4} \nn \,.
\end{align}

\begin{figure}[!b]
\center
\includegraphics[width=0.95\columnwidth]{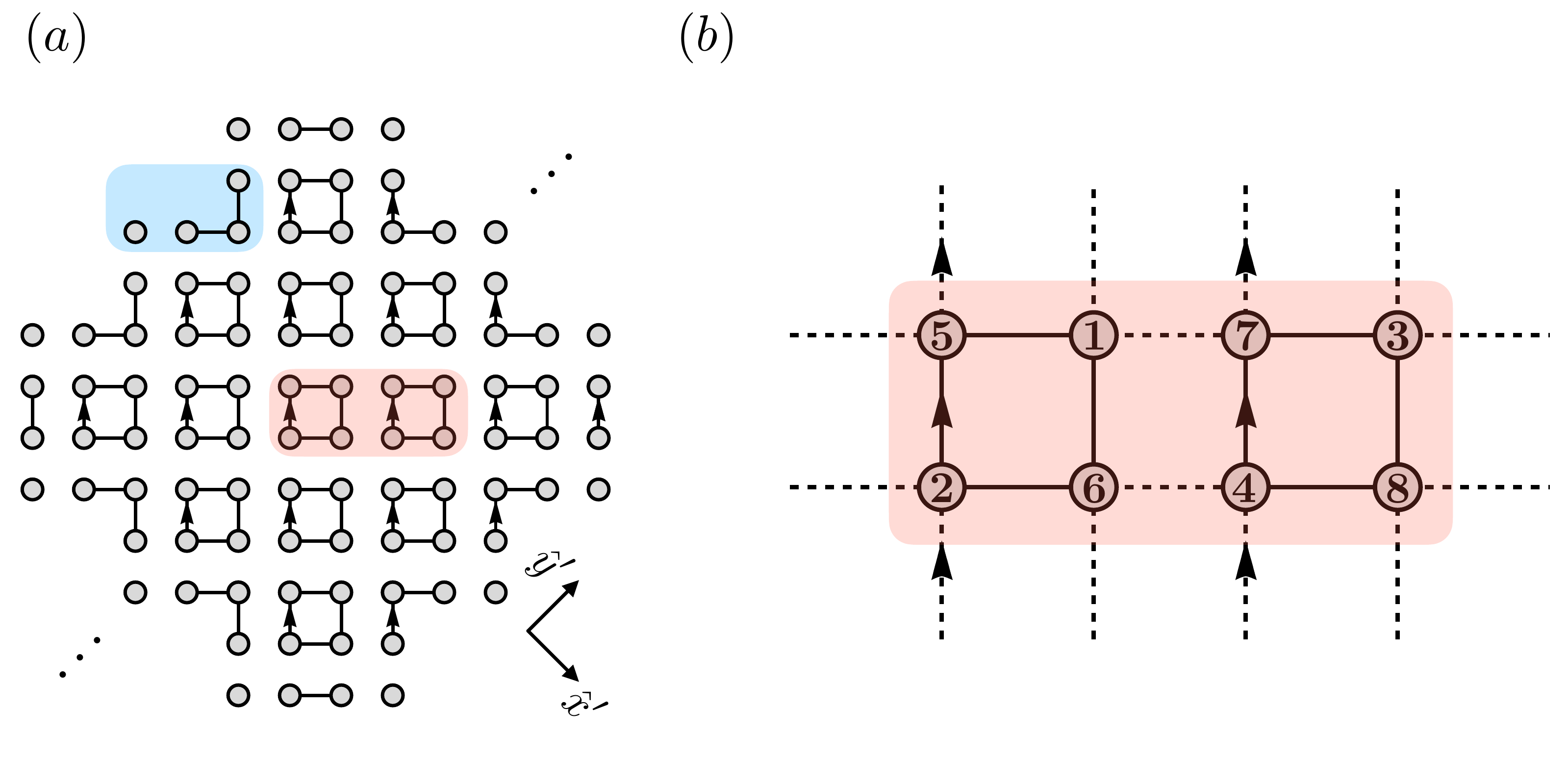}

\caption{Edge states. (a) Stripe geometry with periodic boundary conditions along $\hat y'$. (b) Unit cell choice used to develop the continuum theory. 
}
\label{fig:edge2}
\end{figure}

\textbf{Real-space wavepacket dynamics.}
To validate the semiclassical real-space dynamics, we have numerically simulated the evolution of a real space wavepacket using a finite size $L_x \times L_y$ lattice. We start by constructing a Gaussian wavepacket centered at $\b k_0 = \Gamma = (0,0)$ of the form
\be
|\psi_{\b r,\mu} (t=0) \ket  = \sum_{\b k} e^{- k^2/2\sigma^2} e^{i \b k \cdot (\b r_\mu - \b r_0)} |u_{\b k,\mu} (0) \ket \,,
\ee
where $\mu$ indicates the sublattice degree of freedom in the unit cell and $\b r_\mu$ is its spatial position. In the simulations we take a grid of $k_{pts}\times k_{pts}$ points in $\b k$ space within an interval $\b k \in [-3\sigma, 3\sigma]\times [-3\sigma, 3\sigma]$. We evolve the state with the real-space Hamiltonian $\hat H$ and calculate the observable $\b r^{\rm{num}}(t) = \mv{\psi(t) | \hat{\b r} | \psi(t)}$, with $ \hat{\b r} \equiv \sum_{\b r} (\b r-\b r_0) \, |\b r\ket \bra \b r |$.

We also study the exact evolution of the Bloch wave vector, by considering the velocity operator $\hat {\b v}(\b k) = \pa \hat H(\b k)/\pa \b k = \sum_i \pa_{\b k} d_i(\b k) \Gamma^i$. At each time $t$ the velocity reads $\b v(\b k (t)) = \mv{\psi(\b k(t)) | \hat{\b v} |\psi(\b k(t))}$, where $|\psi(\b k(t))\ket = \mathcal T \exp[-i \int_0^t \rm d t \,\hat H(\b k(t)) ] |u_{\Gamma}(0)\ket $ and $\b k(t) = \b k(0) + \b F t$. This method corresponds to solving the Schr\"odinger equation in $\b k$ space. We find the displacement by integration $\b r^{\rm{exact}}(t) = \int_0^t \rm d t \, \b v(t) + \b r_0$.

\textbf{Edge states.}
Here we derive the effective theory at the edge by considering periodic boundary conditions along the $y'$ direction (see Fig.~\ref{fig:edge2}) for $J_1 \approx J_2$. In this stripe geometry, we have to double the unit cell to correctly represent the lattice periodicity, which reads $d' = 2\sqrt 2 a$. In chiral form, the Hamiltonian reads
\be
H_{\textrm{s}}(\b k) = 
\begin{pmatrix}
0 & Q_s(\b k) \\
Q_s(\b k)^\dag & 0
\end{pmatrix} \,,
\ee
where 
\begin{widetext}
\be
Q_s(\b k) = 
\begin{pmatrix}
-J_1 & -J_1 & -J_2 & -J_2 e^{-i k_x} \\
J_1 & -J_1 & J_2 e^{-i k_y} & - J_2 e^{-i(k_x+k_y)} \\
-J_2 e^{i(k_x+k_y)} & -J_2 e^{i k_y} & -J_1 & -J_1 \\
J_2 e^{ik_x} & -J_2 & J_1 & -J_1 
\end{pmatrix} \,.
\ee
\end{widetext}
Here, we are taking units $d' = 1$ and we are also using the convention that the lattice points within the unit cell are all sitting in the center of the unit cell. The Hamiltonian $H_s(\b k)$ is therefore in Bloch form.

In order to build an effective theory near zero energy \cite{Zhang2009, Shen2012}, let us take $J_2 = (1+m)J_1$, with $|m|\ll 1$. We can then split the Hamiltonian into $\hat H_{\textrm{s}}(\b k) = \hat H_{\textrm{s}}(\b k = 0) + \hat V_{\textrm{s}}(\b k)$, where $\hat V_{\textrm{s}}(\b k)$ is expanded to lowest order in $\b k$. The zeroth order term $\hat H_{\textrm{s}}(\b k = 0)$ can be diagonalized and we find four eigenvectors $|v^i_0\ket$ with energy $E=\pm \sqrt 2 m J_1$, which we use as a basis for the effective theory, and four high-energy states $|v^i_e\ket$ that we neglect. We can then construct the projection operator $\hat P_s = \sum_i |v^i_0\ket \bra v^i_0|$ to obtain at lowest order 
\begin{align}
\hat H_s^{\textrm{eff}}(k_{x'}) &= \hat P_s [\hat H_{\textrm{s}}(\b k = 0) + \hat V_{\textrm{s}}(k_{x'}, k_{y'}=0) ]\hat P_s  \nn\\
&= \begin{pmatrix}
H_\uparrow(k_{x'}) & 0 \\
0 & H_\downarrow(k_{x'})
\end{pmatrix} \,,
\end{align}
where we have rearranged the order of the components to have the Hamiltonian in block-diagonal form and we have defined
\be
H_{\uparrow,\downarrow}(k_{x'}) =  \left( -\f{2m}{\sqrt 2} + \f{1+m}{2\sqrt 2} k_{x'}^2 \right) \sigma_3 \mp \f{1+m}{\sqrt 2} k_{x'} \sigma_2 \,,
\ee
in units where $J_1 = 1$.

We can now substitute $k_{x'} \ra -i \pa_{x'}$ and use $m \ll 1$ to obtain the coupled equations
\be
\left( 2 m + \f{1}{2} \pa_{x'}^2 \right) \sigma_2 \psi(x') = \pm \pa_{x'} \psi(x') \,.
\ee
We use standard procedures to solve these equations, namely we take $\psi(x')$ as an eigenstate of $\sigma_2$, i.e. we decompose it as $\psi(x')=\varphi(x') \chi_\eta$, where $\sigma_2 \chi_\eta = \eta \chi_\eta$ with $\eta = \pm 1$. After taking the ansatz $\varphi(x') \propto e^{-t x'}$, we find that the following algebraic equations must be satisfied

\be
t^2 \pm 2\eta t+ 4m = 0\,,
\ee
for $H_\uparrow(k_{x'})$ and $H_\downarrow(k_{x'})$, respectively. Let us focus on the solution for $H_\uparrow(k_{x'})$, namely the one with plus sign. We find $t_\uparrow=-\eta \pm\sqrt{1-4m} \approx -\eta \pm (1-2m)$. For $\eta = -1$, we can construct a solution $\varphi(x') = c_1 e^{-t^+_\uparrow x'} + c_2 e^{-t^-_\uparrow x'}$ that is exponentially localized for $m>0$ and that vanishes at $x'=0$, namely $c_1 = -c_2$. The solution constructed for $\eta=1$ does not satisfy these requirements for any value of $m$. A similar reasoning can be repeated for $H_\downarrow(k_{x'})$, where we have to take the solution with $\eta=1$ in this case and the solution only exists for $m>0$. We end up with the two zero-energy solutions
\begin{align}
\psi_\uparrow(x') & = 
\begin{pmatrix}
\chi_- \\
0
\end{pmatrix}
e^{-2x'} \left( e^{2mx'} - e^{-2mx'} \right)\,, \nn \\
\psi_\downarrow(x') & = 
\begin{pmatrix}
0 \\
\chi_+
\end{pmatrix}
e^{-2x'} \left( e^{2mx'} - e^{-2mx'} \right) \,,
\end{align}
that are localized at the edge $x'=0$ and that exist for $m>0$, namely for $J_2 > J_1$.


\clearpage
\setcounter{equation}{0}
\setcounter{figure}{0}
\setcounter{section}{0}
\renewcommand{\theequation}{S\arabic{equation}}
\renewcommand{\thesection}{S\arabic{section}}
\renewcommand{\thefigure}{S\arabic{figure}}

\onecolumngrid
\begin{center}
\textbf{
\Large{Supplemental Material: Non-Abelian Bloch oscillations in higher-order topological insulators}
}
\end{center}
\twocolumngrid

\section{Symmetries and winding number}

The BBH model can be casted in the form  
\begin{align}
\hat H(\b k) &= \sum_{i=1}^4 d_i(\b k) \Gamma^i = 
\begin{pmatrix}
0 & Q(\b k) \\
Q(\b k)^\dag & 0
\end{pmatrix}\, \\
Q(\b k) &= d_4(\b k) \mathcal I + i d_i(\b k) \sigma^i \,,
\end{align}
which explicitly shows the chiral symmetry of the model. The doubly degenerate energies are $E = \pm \epsilon_{\b k}$, where $\epsilon_{\b k} = \sqrt{d_1^2 + d_2^2 + d_3^2 + d_4^2}$. Moreover, notice that $Q(\b k)^\dag = \epsilon_{\b k} Q^{-1}$. The lowest two eigenstates can be written as 
\begin{align}
|u_1(\b k) \ket &= \f{1}{\sqrt 2 \epsilon_{\b k}} (d_1(\b k) \!-\! id_2(\b k), -d_3(\b k) \!-\! id_4(\b k), 0, i\epsilon(\b k))^T \,, \nn \\
|u_2(\b k)\ket &= \f{1}{\sqrt 2 \epsilon_{\b k}} (d_3(\b k) \!-\! id_4(\b k), d_1(\b k) \!+\! id_2(\b k), i\epsilon(\b k),0)^T\,, 
\end{align}
that can be compactly written as 
\be
v_\alpha(\b k) = \f{1}{\sqrt 2}
\begin{pmatrix}
- Q(\b k) \xi_\alpha/\epsilon_{\b k} \\
\xi_\alpha
\end{pmatrix}\,,  \quad
\xi_1 = 
\begin{pmatrix}
0 \\ i
\end{pmatrix} , \, 
\xi_2 = 
\begin{pmatrix}
i \\ 0
\end{pmatrix} \,.
\ee

Let us consider the following non-commuting mirror symmetries $\hat M_x = \sigma_1 \otimes \sigma_3$ and $\hat M_y = \sigma_1 \otimes \sigma_1$. Without assuming a specific model we can show that a chiral symmetric Hamiltonian satisfies these mirror symmetries $\hat M_x \hat H(k_x,k_y) \hat M_x^{-1} = \hat H(-k_x,k_y)$ and $\hat M_y \hat H(k_x,k_y) \hat M_y^{-1} = \hat H(k_x,-k_y)$ if and only if
\begin{align}
\label{eq:mx}
d_1(k_x,k_y) & \stackrel{\hat M_x}{=}  +d_1(-k_x, k_y) \,, \nn \\
d_2(k_x,k_y) & \stackrel{\hat M_x}{=}  +d_2(-k_x, k_y) \,, \nn \\
d_3(k_x,k_y) & \stackrel{\hat M_x}{=}  -d_3(-k_x, k_y)\,, \nn \\
d_4(k_x,k_y) & \stackrel{\hat M_x}{=}  +d_4(-k_x, k_y) \,, 
\end{align}
and
\begin{align}
\label{eq:my}
d_1(k_x,k_y) & \stackrel{\hat M_y}{=} -d_1(k_x, -k_y) \,, \nn \\
d_2(k_x,k_y) & \stackrel{\hat M_y}{=}  +d_2(k_x, -k_y) \,, \nn \\
d_3(k_x,k_y) & \stackrel{\hat M_y}{=}  +d_3(k_x, -k_y)\,, \nn \\
d_4(k_x,k_y) & \stackrel{\hat M_y}{=}  +d_4(k_x, -k_y) \,.
\end{align}
We also consider the $\hat C_4$ symmetry, namely $\hat C_4 \hat H(k_x,k_y) \hat C_4^{-1} = \hat H(k_y, -k_x)$, represented by 
\be
\hat C_4 = 
\begin{pmatrix}
0 & \mathcal I \\
-i\sigma_2 & 0
\end{pmatrix}\,.
\ee
This symmetry translates into
\begin{align}
\label{eq:c4}
d_1(k_x,k_y) & \stackrel{\hat C_4}{=}  +d_3(k_y, -k_x) \,, \nn \\
d_2(k_x,k_y) & \stackrel{\hat C_4}{=}  +d_4(k_y, -k_x) \,, \nn \\
d_3(k_x,k_y) & \stackrel{\hat C_4}{=}  -d_1(k_y, -k_x)\,, \nn \\
d_4(k_x,k_y) & \stackrel{\hat C_4}{=}  +d_2(k_y, -k_x) \,.
\end{align}

We will now demonstrate that, along the closed path $\mathcal C$, the previous symmetries quantize the following quantity 
\begin{align}
w_{_\mathcal C} = &\, \f{i}{2\pi} \int_{\mathcal C} \mathrm d \b k\,  \mathrm{Tr} \left[ Q(\b k)^{-1} \sigma_3 \, \pa_{\b k} Q(\b k) \right] \nn\\
= &\,  -\f{1}{\pi}  \int_{\mathcal C} \mathrm d \b k\, \f{1}{\epsilon_{\b k}} \left[ d_1(\b k) \pa_{\b k} d_2(\b k) - d_2(\b k) \pa_{\b k} d_1(\b k) \right. \nn\\
& \, \left. + d_3(\b k) \pa_{\b k} d_4(\b k) - d_4(\b k) \pa_{\b k} d_3(\b k)\right] \,,
\end{align}
and that such a quantity is a winding number. Let us now focus on the path $\mathcal C$ and use the following hypothesis 
\be
\label{eq:d}
d_1 = d_1(k_y)\,, d_2 = d_2(k_y)\,, d_3 = d_3(k_x)\,, d_4 = d_4(k_x)\,,
\ee
namely that the $d_i$ vectors are functions of only one momentum component, which is satisfied by the BBH model. Then the integrand of $w_{_\mathcal C}$ can be written as
\be
w_{_\mathcal C} =  - \f{1}{\pi}  \int_{0}^{2\pi} \f{\mathrm d k}{\epsilon_k^2} w^{(x)}_{_\mathcal C}  - \f{1}{\pi}  \int_{0}^{2\pi} \f{\mathrm d k}{\epsilon_k^2} w^{(y)}_{_\mathcal C} \,.
\ee
\begin{widetext}
We can calculate the two terms separately
\begin{align}
w^{(x)}& = d_1(\b k) \pa_{k_x} d_2(\b k) - d_2(\b k) \pa_{k_x} d_1(\b k) + d_3(\b k) \pa_{k_x} d_4(\b k) - d_4(\b k) \pa_{k_x} d_3(\b k) \nn \\
& \stackrel{\eqref{eq:d}}{=} d_3(k_x, k_y) \pa_{k_x} d_4(k_x, k_y) - d_4(k_x, k_y) \pa_{k_x} d_3(k_x, k_y) \nn \\
& \stackrel{\eqref{eq:c4}}{=} - d_1(k_y, -k_x) \pa_{k_x} d_2(k_y, -k_x) + d_2(k_y, -k_x) \pa_{k_x} d_1(k_y, -k_x) \nn \\
& \stackrel{\eqref{eq:my}}{=} d_1(k_y, k_x) \pa_{k_x} d_2(k_y, k_x) - d_2(k_y, k_x) \pa_{k_x} d_1(k_y, k_x) \nn \\
&  \stackrel{\eqref{eq:d}}{=} d_1(k) \pa_{k} d_2(k) - d_2(k) \pa_{k} d_1(k)\,.
\end{align}
Analogously, for the other term
\begin{align}
w^{(y)}& = d_1(\b k) \pa_{k_y} d_2(\b k) - d_2(\b k) \pa_{k_y} d_1(\b k) + d_3(\b k) \pa_{k_y} d_4(\b k) - d_4(\b k) \pa_{k_y} d_3(\b k) \nn \\
& \stackrel{\eqref{eq:d}}{=} d_1(k_x, k_y) \pa_{k_y} d_2(k_x, k_y) - d_2(k_x, k_y) \pa_{k_y} d_1(k_x, k_y) \nn \\
& = d_1(k) \pa_{k} d_2(k) - d_2(k) \pa_{k} d_1(k) \,.
\end{align}
\end{widetext}
We then find after noticing that $d_1(k) = d_3(k)$ and $d_2(k) = d_4(k)$ (which we justify below based on the combination of $\hat C_4$ and $\hat M_y$ symmetries)
\be
w_{_\mathcal C} =  -\f 1 \pi \int_0^{2\pi} \mathrm d k \, \f{ d_1(k) \pa_{k} d_2(k) - d_2(k) \pa_{k} d_1(k)}{|d_1(k)|^2 + |d_2(k)|^2} \,.
\ee
For the BBH model we obtain
\be
w_{_\mathcal C} =  \mathrm{sign} (J_1^2- J_2^2 ).
\ee
Let us now consider the combination of $\hat C_4$ and $\hat M_y$, namely $\hat M_y C_4 \hat H(k_x,k_y) \hat C_4^{-1} \hat M_y^{-1} = \hat M_y \hat H(k_y, -k_x) \hat M_y^{-1} = \hat H(k_y, k_x)$, which is nothing else than a mirror symmetry with respect to the diagonal axis.  This condition constrains the vectors $d_i$ as follows. Let us consider in particular the set of points $k_x = k_y$. By explicitly calculating the $\hat M_y \hat C_4$ mirror symmetry condition at $k_x = k_y$ for a Dirac Hamiltonian respecting \eqref{eq:d}, we immediately find that $d_1(k) = d_3(k)$ and $d_2(k) = d_4(k)$.

The last task is to connect the winding number with the Wilson loop operator. Let us now consider the Berry connection
\begin{align}
A_x^{12}(\mathcal C) = &\, i \mv{u_1(\b k) | \pa_{k_x} u_2(\b k) }_{\mathcal C} \\
=& \, \f{1}{2 \epsilon_k ^2} \left[ (d_3 - i d_4)\pa_{k_x}(d_2 - i d_1) \right. \nn \\
& \, \left.+ (d_1 + i d_2) \pa_{k_x} (i d_3 + d_4) \right]_{\mathcal C} \nn \\
= &\,  \f{1}{2 \epsilon_k ^2}  \left [(d_1 + i d_2) \pa_{k} (i d_1 + d_2) \right]_{k_x=k_y=k} \nn \\
= &\,  \f{1}{2\epsilon_k ^2} (d_1 \pa_k d_2 - d_2 \pa_k d_1) + \f{i}{2\epsilon_k ^2} (d_1 \pa_k d_1 + d_2 \pa_k d_2) \nn \,.
\end{align}
The $y$ component reads
\begin{align}
A_y^{12}(\mathcal C) = & \, i \mv{u_1(\b k) | \pa_{k_y} u_2(\b k) }_{\mathcal C} \\
= & \, \f{1}{2 \epsilon_k ^2} \left[ (d_3 - i d_4)\pa_{k_y}(d_2 - i d_1) \right. \nn \\
& \, \left. + (d_1 + i d_2) \pa_{k_y} (i d_3 + d_4) \right]_{\mathcal C} \nn \\
=& \, \f{1}{2 \epsilon_k ^2}  (d_1 - i d_2) \pa_{k}(d_2 - i d_1) \nn \\
= & \,  \f{1}{2\epsilon_k ^2} (d_1 \pa_k d_2 - d_2 \pa_k d_1) - \f{i}{2\epsilon_k ^2} (d_1 \pa_k d_1 + d_2 \pa_k d_2) \nn \,.
\end{align}
We therefore find that 
\be
\int_{\mathcal C} ( \mathrm d k_x A^{12}_x +  \mathrm d k_y A^{12}_y ) = \f 1 2 \int_0^{2\pi}\mathrm d k \f{d_1 \pa_k d_2 - d_2 \pa_k d_1}{|d_1|^2 + |d_2|^2} = -\f \pi 2 w_{_\mathcal C}\,.
\ee
The other component of the Berry connection reads
\begin{align}
A_x^{21}(\mathcal C) = & \, i \mv{u_2(\b k) | \pa_{k_x} u_1(\b k) }_{\mathcal C} \\
= & \, \f{1}{2 \epsilon_k ^2} \left[ (d_3 + i d_4)\pa_{k_x}(d_2 + i d_1) \right. \nn \\
& \, \left. + (d_1 - i d_2) \pa_{k_x} (-i d_3 + d_4) \right]_{\mathcal C} \nn \\
= & \,  \f{1}{2 \epsilon_k ^2}  (d_1 - i d_2) \pa_{k_x} (-i d_3 + d_4) \nn \\
= & \, \f{1}{2 \epsilon_k ^2}  (d_1 - i d_2) \pa_{k} (-i d_1 + d_2) \nn \\
= & \, \f{1}{2 \epsilon_k ^2}  (d_1 \pa_k d_2 - d_2 \pa_k d_1) - \f{i}{2 \epsilon_k ^2}  (d_1 \pa_k d_1 + d_2 \pa_k d_2) \nn \,,
\end{align}
whereas 
\begin{align}
A_y^{21}(\mathcal C) = & \, i \mv{u_2(\b k) | \pa_{k_y} u_1(\b k) }_{\mathcal C}\\
= & \, \f{1}{2 \epsilon_k ^2} \left[ (d_3 + i d_4)\pa_{k_y}(d_2 + i d_1) \right. \nn \\
& \, \left. + (d_1 - i d_2) \pa_{k_y} (-i d_3 + d_4) \right]_{\mathcal C} \nn \\
= & \, \f{1}{2 \epsilon_k ^2}  (d_3 + i d_4)\pa_{k_y}(d_2 + i d_1) \nn \\
= & \, \f{1}{2 \epsilon_k ^2}  (d_1 + i d_2)\pa_{k_y}(d_2 + i d_1)\nn \\
= & \, \f{1}{2 \epsilon_k ^2}  (d_1 \pa_k d_2 - d_2 \pa_k d_1) + \f{i}{2 \epsilon_k ^2}  (d_1 \pa_k d_1 + d_2 \pa_k d_2) \nn \,,
\end{align}
and we finally conclude that 
\be
\int_{\mathcal C} ( \mathrm d k_x A^{21}_x +  \mathrm d k_y A^{21}_y ) =  -\f \pi 2 w_{_\mathcal C}\,.
\ee
Moreover, notice that $A^{12}_i(\b k) = [A^{21}_i(\b k)]^*$ as required by SU(2).

Let us now have a look at the diagonal components of the Berry connection
\begin{align}
A^{11}_x(\mathcal C) &=  \f{1}{2 \epsilon_k ^2} \left[ d_1 \pa_{k_x} d_2 - d_2 \pa_{k_x} d_1 - d_3 \pa_{k_x} d_4 + d_4 \pa_{k_x} d_3 \right]_{\mathcal C} \nn \\
&= \f{1}{2 \epsilon_k ^2} ( - d_3 \pa_{k} d_4 + d_4 \pa_{k} d_3)  \nn \\
&=  \f{1}{2 \epsilon_k ^2} ( - d_1 \pa_{k} d_2 + d_2 \pa_{k} d_1)\,,
\end{align}
whereas
\begin{align}
A^{11}_y(\mathcal C) &=  \f{1}{2 \epsilon_k ^2} \left[ d_1 \pa_{k_y} d_2 - d_2 \pa_{k_y} d_1 - d_3 \pa_{k_x} d_4 + d_4 \pa_{k_x} d_3 \right]_{\mathcal C} \nn \\
&= \f{1}{2 \epsilon_k ^2} ( d_1 \pa_{k} d_2 - d_2 \pa_{k} d_1)
\end{align}
thus concluding that $A^{11}_x(\mathcal C)+A^{11}_y(\mathcal C)=0$ which shows that the Wilson loop on the path $\mathcal C$ is only off-diagonal, and in particulare that 
\be
\int_\mathcal{C} \mathrm d \b k \cdot \b A(\b k) = \pm \f \pi 2 \sigma_1 \,.
\ee

By using crystal symmetries and the combination $\hat C_4 \hat M_x$, similar relations can be obtained for the $\bar{\mathcal C}$ path, where we find that 
\be
\int_{\bar{\mathcal{C}}} \mathrm d \b k \cdot \b A(\b k) = \pm \f \pi 2 \sigma_3 \,.
\ee

\begin{figure}[!t]
\center
\includegraphics[width=0.9\columnwidth]{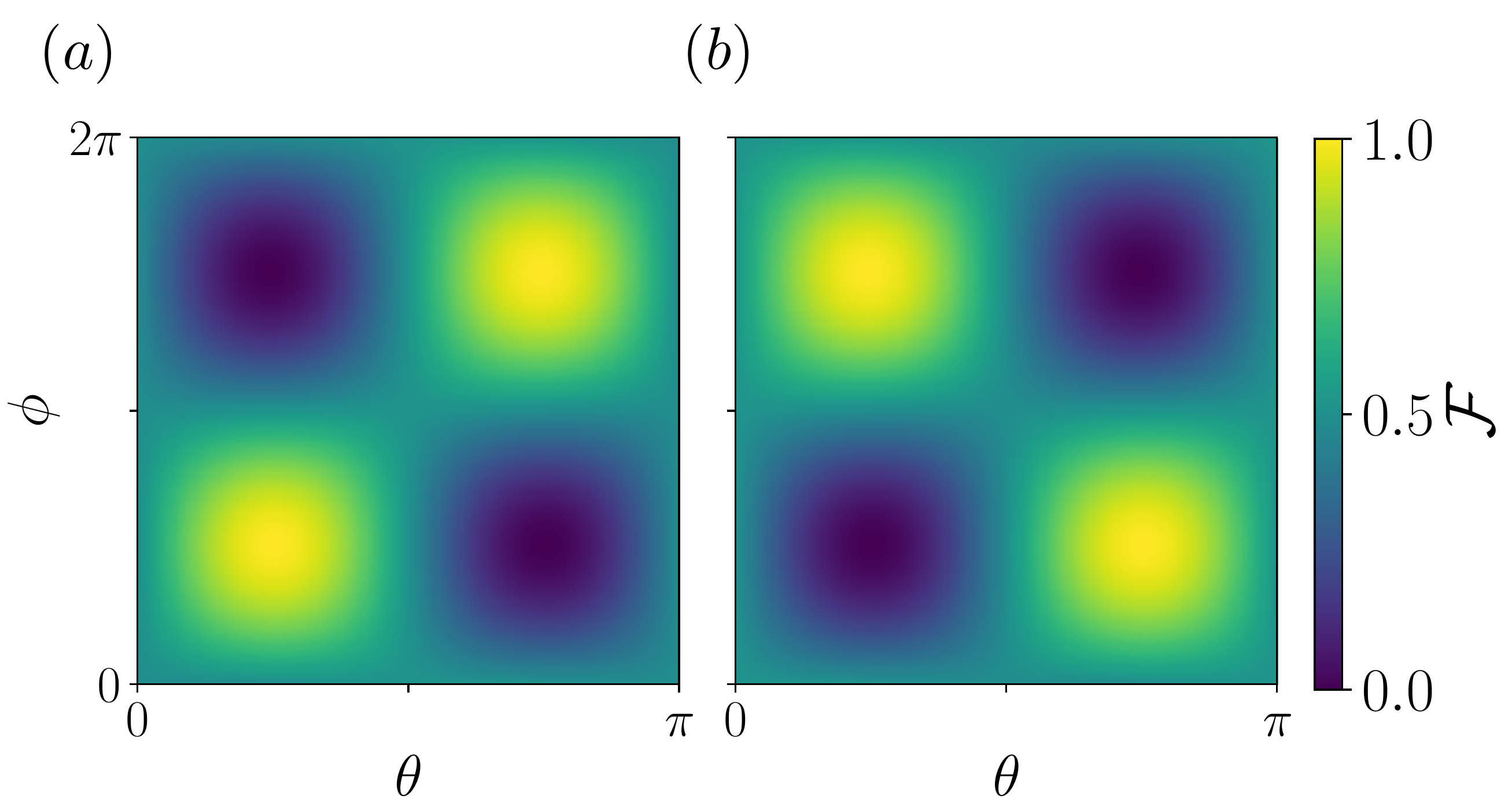}

\caption{Fidelity of (a) the ground state $\mathcal F^1$ and (b) of the excited state $\mathcal F^2$ as a function of the BBH eigenstates degenerate manifold for flux $\varphi=\pi-0.1$ and $J_2 = 0.5 J_1$.  
}
\label{fig:fidelity}
\end{figure}

\section{Degeneracy breaking}

Here, we consider a state preparation protocol based on the breaking of time-reversal symmetry in the BBH model. Let us consider the vertical hopping coefficients responsible for the $\pi$ flux to have a generic complex dependence $e^{i \varphi}$, corresponding to plaquettes with staggered flux $\pm \varphi$. The eigenstates at the $\Gamma$ point read
\begin{widetext}
\begin{align}
|u_\Gamma^1(\varphi) \ket &= \f{1}{2} \left( |\sin(\varphi/4)| (1+i\cot(\varphi/4)), \f{4|\sin(\varphi/4)|\cos(\varphi/2)}{2\cos(\varphi/2)-\cos \varphi +i \sin \varphi - 1}, -\cos(\varphi/2)+i\sin(\varphi/2), 1 \right)^T \,, \\
|u_\Gamma^2(\varphi) \ket &= \f{1}{2} \left( |\cos(\varphi/4)| (1-i \tan(\varphi/4)), \f{4|\cos(\varphi/4)|\cos(\varphi/2)}{2\cos(\varphi/2)+\cos \varphi - i \sin \varphi + 1},  \cos(\varphi/2)-i\sin(\varphi/2), 1 \right)^T \,, \nn
\end{align}
\end{widetext}
with energies $E_1= -(J_1 + J_2) |\sin(\varphi/4)|$ and $E_2 = -(J_1 + J_2) |\cos(\varphi/4)|$. Let us now consider a generic combination of the $\pi$ flux eigenstates as considered in the main text, namely $|u_\Gamma(\theta,\phi)\ket = \cos(\theta)|u^1_\Gamma\ket + \sin(\theta) e^{i\phi}|u^2_\Gamma\ket$. For $\varphi = \pi-0.1$, the lowest energy state is $|u_\Gamma^1(\varphi)\ket$ and we can then calculate the fidelity $\mathcal F^\alpha = |\bra u_\Gamma(\theta,\phi) |u_\Gamma^\alpha(\varphi)\ket|^2$, which is shown in Fig.~\ref{fig:fidelity}. We therefore find that the ground state is a distribution of the degenerate BBH eigenstates peaked at $\theta = \pi/4, \, 3\pi/4$ and $\phi = \pi/2,\, 3\pi/2$. An analogous reasoning can be repeated for the excited state $|u_\Gamma^2(\varphi) \ket$.

A protocol for state preparation would then require to slightly break time-reversal symmetry in order to prepare a BEC occupying the ground state $|u_\Gamma^1(\varphi) \ket$. Then, one can treat the states $|u_\Gamma^{1,2}(\varphi) \ket$ as a two-level system and apply a coherent external coupling with frequency $\omega = \Delta E = E_2-E_1$ to make a superposition of $|u_\Gamma^1(\varphi) \ket$ and $|u_\Gamma^2(\varphi) \ket$ with relative imbalance (parametrized by $\theta$) and phase (parametrized by $\phi$) as the initial states discussed in the main text. In order to reproduce the BOs results discussed in this work, the applied force must then satisfy $|F|\gg \Delta E$ such that the two bands are effectively degenerate on the time-scale of the BO.

\end{document}